\documentclass[prb,twocolumn,showpacs,preprintnumbers,amsmath,amssymb]{revtex4-1}
\makeatletter
\def\@dotsep{4.5}
\makeatother
\usepackage[dvips]{graphicx}
\usepackage{amsmath}
\usepackage{amsbsy}
\usepackage{color}
\usepackage{epstopdf}
\DeclareGraphicsExtensions{.pdf,.eps,.png,.jpg,.mps}

\definecolor{textcolor}{cmyk}{0,0,0,1}
\definecolor{magenta}{rgb}{1,0,1}
\definecolor{green}{rgb}{0,1,0}
\definecolor{red}{rgb}{1,0,0}

\begin{document}

\title{ Electronic transport through bilayer graphene flakes}
\author{J. W. Gonz\'alez$^1$, H. Santos$^2$, M. Pacheco$^1$, L. Chico$^2$ and L. Brey$^2$}
\affiliation{$^1$Departamento de F\'{i}sica, Universidad T\'{e}cnica Federico Santa Mar\'{\i}a, Casilla postal 110 V, Valpara\'{i}so, Chile
\\
$^2$Departamento de Teor\'{\i}a y Simulaci\'on de Materiales,
Instituto de Ciencia de Materiales de Madrid, Consejo Superior de
Investigaciones Cient{\'{\i}}ficas, Cantoblanco, 28049 Madrid, Spain}

\date{\today}

\begin{abstract}
We investigate the electronic transport properties of a bilayer graphene flake contacted by two monolayer nanoribbons.
Such a finite-size bilayer flake can be built by overlapping two semi-infinite ribbons or by depositing a monolayer flake onto an infinite
nanoribbon. These two structures have a complementary behavior, that we study and analyze by means of a tight-binding method and
a continuum Dirac model. We have found that for certain energy ranges and geometries, the conductance of these systems oscillates markedly between zero and the maximum value of the conductance, allowing for the design of electromechanical switches.
Our understanding of the electronic transmission through bilayer flakes may provide a way to measure the interlayer hopping in bilayer graphene.
\end{abstract}

\maketitle

\section{\label{sec:intro} Introduction}
Graphene is a sheet of carbon atoms that order in a honeycomb
structure, which is composed of two inequivalent triangular sublattices $A$ and $B$.
Since its experimental
isolation in 2004 \cite{Novoselov_2004} and the subsequent
verification of its exotic properties, the interest in this material
has boosted. Carriers in monolayer graphene behave as
two-dimensional (2D) massless Dirac fermions, \cite{Castro_Neto_RMP}
with a linear dispersion relation $\varepsilon
(\textbf{k})= \pm v_F k$. Phenomena of fundamental nature, such as
quantum Hall effect \cite{Novoselov_2005,Zhang_2005} and
Klein \cite{Young_2009,Stander_2009} tunneling have been recently
measured in graphene based devices.

Being a material of atomic thickness, graphene is regarded as a
promising candidate for nanoelectronic applications. 
\cite{Castro_Neto_RMP} By patterning graphene, its electronic
structure can be altered in a dramatic fashion: size quantization
yields ribbons with electronic gaps, essential for electronics. 
 \cite{Nakada_1996,Fujita_1996,Wakabayashi_1999} By imposing
appropriate boundary conditions, the  physics of graphene
nanoribbons is well described within a continuum Dirac model.
 \cite{Brey_2006a,Brey_2006b,Akhmerov_2008} Furthermore, connections
and devices can be designed in a planar geometry by cutting graphene
layers. \cite{Iyengar_2008} Another way to modify the band structure
of graphene is to stack two graphene monolayers,  $1$ and $2$,
forming a bilayer
graphene. \cite{McCann_2006,Guinea_2006,Castro_2007} In bilayer
graphene there are four atoms per unit cell, with inequivalent sites
$A1$, $B1$ and $A2$, $B2$ in the first and second graphene layers,
respectively.

Different stacking orders can occur in bilayer graphene. Due to its
larger stability for bulk graphite, the most commonly studied is
$AB$ (Bernal) stacking. In the $AB$ stacking, the two graphene
layers are arranged in such a way that the $A1$ sublattice is
exactly on top of the sublattice $B2$. In the simple hexagonal or 
$AA$ stacking, both sublattices of sheet $1$, $A1$ and $B1$, are
located directly on top of the two sublattices $A2$ and $B2$ of sheet
$2$. Although graphite with direct or $AA$ stacking has not been observed in natural graphite, 
it has been produced by folding graphite layers at the edges of a cleaved sample with a scanning tunneling microscope tip; \cite{Roy_1998}
additionally, the growth of $AA$-stacked graphite on (111) diamond has also been reported. \cite{Lee_2008} 
Furthermore, it has been recently found that $AA$ stacking is
surprisingly frequent in bilayer graphene, \cite{Liu_2009} so it
should be also considered as a realistic possibility in few-layer
graphene. The interplanar spacing for the $AB$ stacking has been experimentally determined to be 
 $c_{AB} = 3.35$\AA , \cite{Hanfland_1989}, whereas for the $AA$ stacking seems to be somewhat larger, $c_{AA} \sim 3.55$\AA . \cite{Lee_2008}
 First-principles calculations agree with these values. \cite{Charlier_1994,Palser_1999,Xu_2010}
 In any case, the distance
between atoms belonging to different layers in both stackings is much
larger that the separation between atoms in the same layer, $a_{CC} = 
1.42$ \AA .

Nanostructures based on bilayer graphene have begun to be explored
only recently. \cite{Min_2009,Nilsson_2007,Castro_2007,Snyman_2007,Fiori_2009a,Fiori_2009b}
Bilayer graphene nanoribbons might present better
signal-to-noise ratio in transport experiments than monolayer
ribbons. \cite{Lin_2008} Graphene flakes are quantum-dot-like
structures, and because of their aspect ratio they are also called
nanobars. Both, bilayer nanoribbons and bilayer flakes, show
interesting properties with an intriguing dependence on stacking.
The dependence of the energy gap of bilayer graphene flakes on their
width and length as well as on their atomic termination has been
recently reported. \cite{Sahu_2009}

\begin{figure}[ht]
\includegraphics[width=\columnwidth,clip]{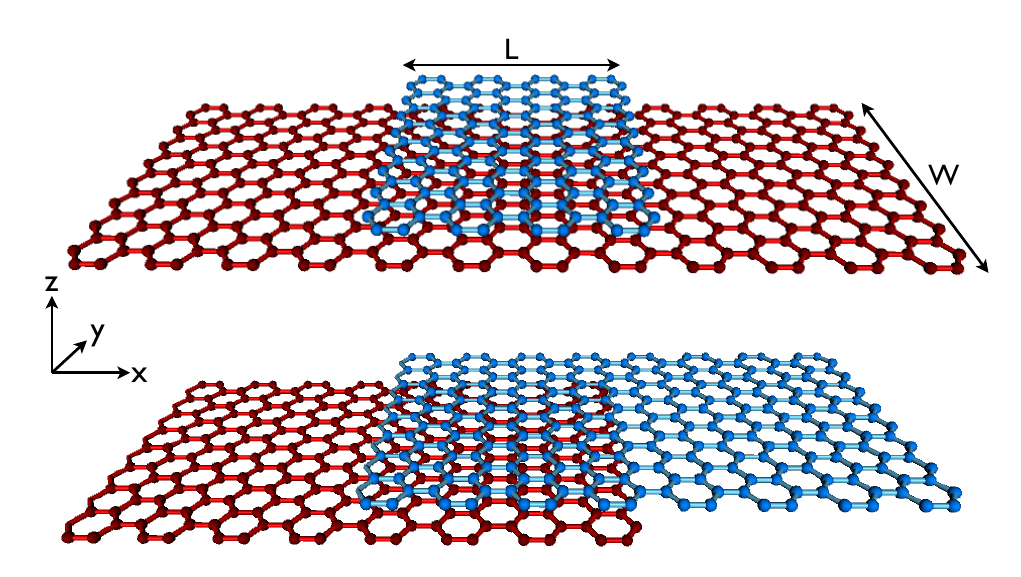}
\caption{(color online) Schematic view of two possible geometries
for a bilayer graphene flake contacted by two nanoribbons. 
Top: A finite-size bilayer
graphene flake achieved by overlaying a monolayer graphene quantum
dot over an infinite graphene nanoribbon ($1\rightarrow 1$ configuration).
Bottom: The
bilayer graphene flake is formed by the overlap of two semi-infinite
nanoribbons ($1 \rightarrow 2$ configuration). 
In both cases the width and length of the bilayer region are $L$ and
$W$ respectively.}
\label{fig:flakes}
\end{figure}

In this paper we concentrate in the transport properties of bilayer 
armchair 
graphene flakes with nanoribbon contacts. We consider that the
most likely way of achieving such quasi-zero dimensional structures
is either by the overlap of two nanoribbons, depicted in the lower part of Fig.
\ref{fig:flakes}, or the deposition of a finite-size graphene
flake over a graphene nanoribbon, shown in the upper part of Fig. \ref{fig:flakes}. 
These two configurations correspond to two different ways
of providing monolayer nanoribbon leads for the bilayer flake:
either the ribbon leads are contacted to different layers of the
flake, or to the same
monolayer. We will address
these two configurations as bottom-bottom ($1 \rightarrow 1$) or
bottom-top ($1 \rightarrow 2$), respectively. In both geometries the
width of the bilayer flake and nanoribbons is the same, $W$, and
the length of the bilayer region is $L$. In this
work we consider narrow 
armchair metallic graphene nanoribbons 
in the energy range for which only one incident
electronic channel is active.

We calculate the conductance with two different approaches: a
tight-binding model using the Landauer-B\"uttiker formalism and a
mode-matching calculation in the continuum Dirac-like Hamiltonian
approximation. Our main results are the following:
\par \noindent
i) In the $AA$ stacking configuration, the transmission through the system
shows antiresonances due to the interference of the two propagating electronic channels in the bilayer flake. 
For a bilayer region of length $L$ we obtain that
the conductance oscillates  as function of energy with a main period $v_F\pi/L$. For a fixed
incident energy $E$, the conductance 
as a function of the length $L$ oscillates with two main periods: $\pi v_F/ \gamma_1 $ and $\pi v_F
/ E$, being $\gamma _1$ the interlayer hopping parameter. The
bonding/antibonding character of the bilayer bands in the $AA$
stacking makes the bottom-top
 and bottom-bottom conductances
to be rather complementary: the conductance is zero in
the bottom-top configuration and it is finite in the
bottom-bottom arrangement at zero energy, and in general the maxima of the bottom-top configuration coincide with the minima 
of the bottom-bottom one and viceversa.

\par \noindent
ii) For the $AB$ stacking, and for energies larger than the
interplane hopping $\gamma_1$, 
these devices behave similar to those in
the $AA$ configuration because there are also two propagating electron
channels in the bilayer flake at these energies. 
The conductance presents
antiresonances with periods depending on $E$, $\gamma _1$, and $L$.
An interesting difference is that for a fixed incident energy, the
period related with interlayer hopping is twice than that found for the $AA$ 
stacking. This reflects that in the $AB$ stacking only
half of the atoms are connected by interlayer hopping, whereas in the $AA$ arrangement all atoms are connected.
\par \noindent
iii) For energies smaller than the interplane hopping, 
for which the $AB$ stacking 
has only a propagating channel, the
conductance shows Fabry-Perot-like resonances. These are associated with
constructive interferences in the only available electronic channel.
At zero energy the conductance of the bottom-bottom configuration is
unity, whereas in the bottom-top geometry the conductance is
zero.

We have analyzed the dependence of the transmission with the
structural parameters and the interlayer coupling in bilayer
graphene. This study provides a way to determine the interlayer
hopping by studying the variation of the low energy conductance of
two overlapping nanoribbons with the bilayer flake length; in
addition, it could clarify the role of stacking in the
transport characteristics of these systems.
Our results also indicate that the conductance, as function of energy and
system size, oscillates markedly between zero and a finite value,
allowing for the design of electromechanical switches based on
overlapping nanoribbons. 
The introduction of an external gate voltage is of interest 
for potential applications, however, we restrict ourselves to zero gate voltage in order to 
obtain analytical expressions in the Dirac model and acquire a physical understanding of the 
transport properties of these structures.

This work is organized as follows.  In Section
\ref{sec:hamiltonian} we introduce the tight-binding and Dirac
Hamiltonians we use to model the electronic properties of
graphene. Section \ref{Conductance} is dedicated to describe the
conductance calculations, both in the tight-binding approximation, for
which we use  Landauer-B\"uttiker formalism, and in the continuum Dirac-like model, 
 where we use a wavefunction matching technique.
Section IV is dedicated to present numerical results obtained in the
tight-binding Hamiltonian and compare them with the analytical
results obtained in the Dirac formalism. Finally, we conclude in Section V summarizing our main results. 

\section{\label{sec:hamiltonian} Theoretical description of the system}

The low energy properties in graphene are mainly determined by the
$p_z$ orbitals. 
Thus, 
we adopt a $\pi$-band tight-binding Hamiltonian with
nearest-neighbor in-plane interaction given by the hopping
parameter $\gamma _0=2.66$ eV. In undoped graphene, the conduction
and valence bands touch at two inequivalent points of the Brillouin zone
$\textbf{K}$ and $\textbf{K}'$. Near these points, the electric
properties of graphene can be described by a massless Dirac
Hamiltonian \cite{Castro_Neto_RMP} that has a linear dispersion
with slope $v_F$=$\frac {\sqrt{3}} 2\gamma_0 a_0$, where $a_0=2.46$ \AA \ is the
graphene in-plane lattice parameter.

Bilayer graphene consists of two graphene layer coupled by tunneling.
The interlayer coupling is modeled with a single hopping $\gamma_1$ connecting 
atoms directly on top of each other, which we take as
$\gamma_1 = 0.1\gamma_0$, in agreement with experimental results.
\cite{Ohta_2006,Malard_2007} As discussed in the Introduction, the interlayer hopping is considerably 
smaller than the intralayer hopping because the nearest-neighbor distance between
carbon atoms is much smaller than the interlayer
separation. 
We do not include other remote terms, such as trigonal warping $\gamma_3$, because even though 
it has a similar value to $\gamma_1$, its effects are more important away from the neutrality point, 
where the Dirac cones are distorted and therefore the continuum approximation is not so good. 

\subsection{Tight-binding Hamiltonians}

The tight-binding Hamiltonian for the $AB$-stacked bilayer reads
\begin{eqnarray}
H^{AB}=& - & \gamma _0 \sum _{<i,j>,m} (a^+_{m,i} b_{m,j} +
h.c.)
\nonumber \\
& - & \gamma_1 \sum _{i} (a^+_{1,i} b_{2,i} + h.c.), \label{H_TB_AB}
\end{eqnarray}
where $a_{m,i} (b_{m,i})$ annihilates an electron on sublattice $A
(B)$, in plane $m=1,2$, at lattice site $i$. The subscript $<i,j>$
represents a pair of in-plane nearest neighbors. For the $AB$ 
stacking we assume that the atoms on the $A$ sublattice of the
bottom layer ($A1$) are connected to those on the $B$ sublattice of
the top layer ($B2$). The second term in Eq. (\ref{H_TB_AB}) 
represents the hopping between these two sets of atoms.

For the bilayer with $AA$ stacking, all the atoms of layer $1$ are on top of the
equivalent atoms of layer $2$; thus, the Hamiltonian takes the form 
\begin{eqnarray}
H^{AA}=& - & \gamma _0 \sum _{<i,j>,m} (a^+_{m,i} b_{m,j} +
h.c.)
\nonumber \\
& - &\gamma_1 \sum _{i} (a^+_{1,i} a_{2,i} + b^+_{1,i} b_{2,i}+h.c.). 
\label{H_TB_AA}
\end{eqnarray}

\begin{figure}[ht]
\includegraphics[width=\columnwidth,clip]{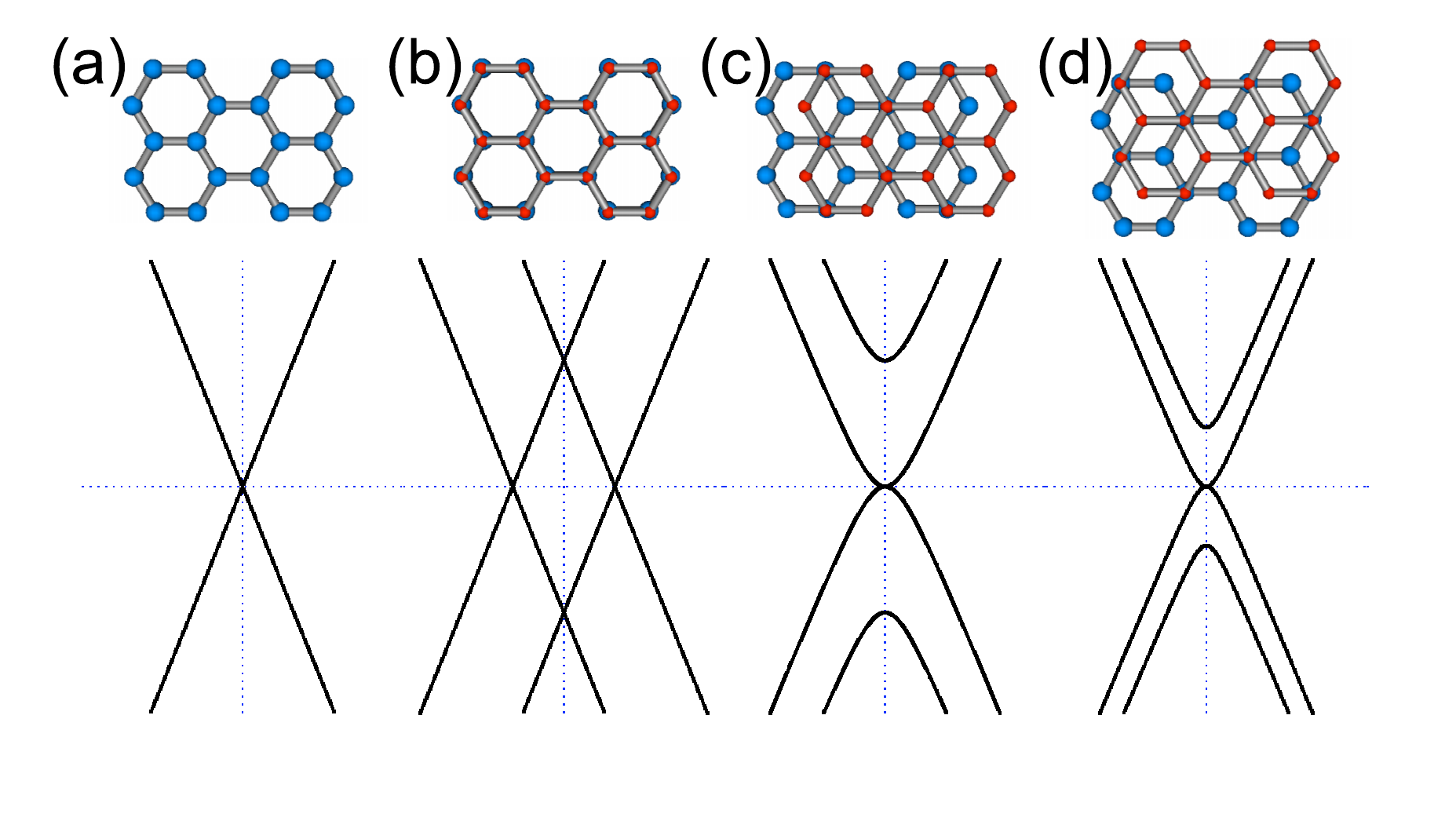}
\caption{(color online) Atomic structure geometries and band
dispersion relations around the Dirac point for several
armchair-terminated nanoribbons. The ribbon longitudinal axes are in
the horizontal direction.  (a) Monolayer armchair nanoribbon; (b)
bilayer nanoribbon with $AA$ stacking; (c) bilayer ribbon
with $AB$-$\alpha$ stacking; (d) bilayer nanoribbon with
$AB$-$\beta$ stacking. For this energy range, the dispersion relations (a)-(c) are independent of the ribbon width; 
case (d) corresponds to $N=17$.}
\label{fig:bands}
\end{figure}

As we are interested in the transport properties of the bilayer
flakes, we will concentrate on structures where the leads are
monolayer armchair graphene nanoribbons (aGNR), with widths chosen
to have metallic character. We denote the ribbon width with an
integer $N$ indicating the number of carbon dimers along it. With
this convention, a nanoribbon of width $N= 3p +2$, where $p= 0, 1, 2
...$, is metallic. In Fig. \ref{fig:bands}(a) we plot the atomic
geometry of the monolayer aGNR leads and the corresponding low
energy electronic bands, as obtained from the tight-binding
Hamiltonian. Note that in aGNR the two Dirac points collapse in just
one. \cite{Brey_2006a} Near the Dirac point the dispersion is linear, $v_F k$.
In the transport calculations we will only
consider incident electrons inside this subband, i.e., with energy
lower than the second subband.
An aGNR is metallic because of a particular combination of the
wavefunctions coming form the two original Dirac points. This
combination is preserved when piling up two metallic armchair
monolayer ribbons, being the corresponding bilayer nanoribbon also
metallic.

 The details of the low energy spectrum of bilayer nanoribbons
depend on the particular stacking. In Fig. \ref{fig:bands}(b) we
plot the tight-binding band structure of a bilayer nanorribon with $AA$ stacking. 
The bands also present a linear dispersion and
they can be understood as bonding/antibonding combinations of the
constituent monolayer aGNR bands.

 The $AB$ stacking can be
achieved from the $AA$ bilayer geometry by displacing one graphene monolayer with respect to the
other, in such a way that the atoms of one sublattice (i.e., $A$) of
the top monolayer are placed over the atoms of the other sublattice
($B$) of the bottom monolayer. In nanoribbons, two different $AB$ 
stackings are possible:\cite{Sahu_2008} the $AB$-$\alpha$ stacking, shown in Fig.
\ref{fig:bands}(c), which yields a more symmetric geometry for
infinite armchair nanoribbons, and the $AB$-$\beta$ stacking,
shown in \ref{fig:bands}(d). Notice that, for armchair nanoribbons,
the $AB$-$\alpha$ configuration can be reached by displacing the top
monolayer in the direct stacking a distance equal to the
carbon-carbon bond $a_{CC}$ along the ribbon length, as can be seen
by comparing Figs. \ref{fig:bands}(b) and (c). For the
$AB$-$\beta$ stacking, the displacement is of the same magnitude
but at 60$^{\rm o}$ with the ribbon longitudinal direction, yielding
a less symmetric configuration for armchair nanoribbons (Fig.
\ref{fig:bands}(d)). In both cases the $AB$-stacked bilayer
graphene nanoribbons have metallic character, and 
the conduction and valence bands have a parabolic dispersion at the Dirac point.

\subsection{\label{sec:dirac}Dirac-like Hamiltonians}

Most of the low energy properties of monolayer and bilayer graphene
nanoribbons can be understood using a  $\mathbf{k} \cdot \mathbf{p} $ 
approximation, which yields a Dirac-like Hamiltonian. \cite{Brey_2006a,Brey_2006b,Nilsson_2007} The 
low-energy effective bilayer Hamiltonian describing the properties of a
infinite   $AA$-stacked bilayer has the form 
\begin{equation}\label{HAA}
  H_{AA} =  \left( {\begin{array}{*{20}{cccc}}
  0 & v_F {{\pi ^ \dag }} & \gamma _1 & 0  \\
   v_F \pi  & 0 & 0 & \gamma _1  \\
  \gamma _1 & 0 & 0 &  v_F {{\pi ^ \dag }}  \\
  0 & \gamma _1 &  v_F \pi  & 0  \\
\end{array} } \right) \, \, \, ,
\end{equation}
where  $\pi = {k_x} +i {k_y} = k e^{i\theta _k }$, $ \theta
_\mathbf{k} = \tan^{ - 1} \left( {k_x } / {k_y } \right)$, and
$\textbf{k} = (k_x,k_y)$ is the momentum relative to the Dirac
point.  The Hamiltonian acts on a four-component spinor $( \phi_A
^{(1)}, \phi_B ^{(1)},\phi_A ^{(2)},\phi_B ^{(2)})$. The
eigenfunctions of this Hamiltonian are bonding and antibonding
combinations of the isolated graphene sheet solutions,
\begin{equation}\label{AutoAA}
\begin{gathered}
\varepsilon _{s,\pm}^{AA}  =s v_F k \pm  \gamma _1 \, , \, \, \; \psi _{s,\pm} ^{AA} = \left( {\begin{array}{*{20}c}
  1  \\
  {s e^{i\theta _{\textbf{k}} } }  \\
  \pm 1  \\
  \pm  s {e^{i\theta _{\textbf{k}} } }
\end{array} } \right)e^{i\mathbf{k}\cdot\mathbf{r}} , \hfill \\
\end{gathered}
\end{equation}
with $s = \pm 1$.

The low-energy Hamiltonian of the $AB$ stacking 
reads\cite{McCann_2006}
\begin{equation}\label{HAB}
H_{AB} =  \left( {\begin{array}{*{20}{cccc}}
  0 &  v_F {{\pi ^ \dag }} & 0 & \gamma _1  \\
   v_F \pi  & 0 & 0 & 0  \\
  0 & 0 & 0 &  v_F {{\pi ^ \dag }}  \\
  \gamma _1 & 0 &  v_F \pi  & 0  \\
\end{array} } \right),
\end{equation}
with eigenvalues
\begin{equation}\label{AutoAB_1}
\varepsilon _{s,\pm}^{AB}  = \frac{s} {2}\left( {  \gamma_1 \pm
\sqrt {4 v_F ^2 k^2  + \gamma _1
^2 } } \right)\, \, \, , s = \pm 1.\, \, \, 
\end{equation}
For a given eigenvalue $E$,the wavefunction takes the form
\begin{equation}\label{AutoAB_2}
\begin{gathered}
\psi _{s,\pm} ^{AB} = \left( {\begin{array}{*{20}c}
  E  \\
  {v_F k e^{i \theta  } }  \\
  - \frac {v_F k e ^{-i\theta}} {\gamma_1 E} (v_F ^2k^2 -E^2)  \\
  -\frac { v_F ^2 k^2 - E^2}{\gamma_1}
\end{array} } \right)e^{i\mathbf{k}\cdot\mathbf{r}} . \hfill \\
\end{gathered}
\end{equation}

In accordance with the geometry shown in Fig. \ref{fig:flakes},  we assume for nanoribbons that the system is invariant in the $x$
direction, and therefore $k_x$ is a good quantum number. In the case
of metallic  aGNR, the boundary conditions are
satisfied\cite{Brey_2006a} for $k_y=$0 independently of the
nanoribbon width; this  $k_y=$0 state is the lowest energy band confined
in the aGNR. 
We have checked that the dispersion of the lowest energy
band obtained by solving the Dirac model coincides with that 
obtained by diagonalizing the tight-binding Hamiltonian for the monolayer, bilayer $AA$ and $AB$-$\alpha$ nanoribbons.
 Therefore, the
Dirac approximation is a good description for the low
energy properties of these nanoribbons, Fig. \ref{fig:bands}(a)-(c). This
is not the situation for bilayer graphene nanoribbons with $AB$-$\beta$
stacking. In this case, the atomic asymmetry at the edges of the
ribbon is not captured by the Dirac model, which is a long-wavelength
approximation. Therefore,  we should describe
the electronic properties of nanoribbons with $AB$-$\beta$ stacking using the tight-binding Hamiltonian.

\section{\label{Conductance} Conductance}
\subsection{Tight-Binding approach: Landauer-B\"uttiker formalism}
Due to the lack of translational invariance of the system, in the
tight binding model we calculate the electronic and transport
properties  using the surface Green function matching method.\cite{Chico_1996b,Nardelli_1999} 
To this end, the system is
partitioned in three blocks: two leads, which we
assume to be semi-infinite aGNR, and the conductor, consisting of the
bilayer flake. The Hamiltonian is
\begin{equation}
H= H_C  + H_R + H_L + V_{LC} + V_{RC},
\end{equation}
where $H_C$, $H_L$, and $H_R$ are the Hamiltonians of the central
portion, left and right leads respectively, and $V_{LC}$, $V_{RC}$
are the coupling matrix elements from the left $L$ and right $R$
lead to the central region $C$. The Green function of the conductor
is
\begin{equation}
\mathcal{G}_C(E) = (E-H_C - \Sigma_L -\Sigma_R)^{-1},
\end{equation}
where $\Sigma_\ell= V_{\ell C}g_\ell V_{\ell C}^\dagger$ is the
selfenergy due to lead $\ell=L,R$, and $g_\ell = (E -H_\ell)^{-1}$
is the Green function of the semi-infinite lead $\ell$.
\cite{Rosales_2008}

In the linear response regime, the conductance can be calculated
within the Landauer formalism as a function of the energy $E$. In
terms of the Green function of the system,\cite{Datta_book,Chico_1996b,Nardelli_1999} it reads
\begin{equation}\label{LandauerG}
G = \frac{{2e^2 }}{h}T\left( {E } \right) = \frac{{2e^2 }}{h}
{\mathop{\rm Tr}\nolimits} \left[ {\Gamma _L \mathcal{G}_C \Gamma _R
\mathcal{G}_C^\dagger} \right],
\end{equation}
where  $T\left( {E } \right)$, is the transmission function across
the conductor, and  $\Gamma_{\ell}=i[ {\Sigma _{\ell}  - \Sigma
_{\ell} ^{\dag} }]$ is the coupling between the conductor and the
$\ell=L,R$ lead.

\subsection{Continuous approximation: wavefunction matching}

In the low-energy limit, we can obtain the conductance of the
system by matching the eigenfunctions of the Dirac-like
Hamiltonians. As commented above, we consider incident electrons from the
lowest energy subband, which correspond to a transversal
momentum $k_y$=0 in aGNRs. Assuming an electron with energy $E$ coming
from the left monolayer ribbon, we compute the transmission
coefficient $t$ to the right monolayer lead. In the central part the
wavefunctions are linear combinations of the solutions of the
bilayer nanoribbon Hamiltonians given in Sec. \ref{sec:dirac} 
at the incoming energy $E$. The transmission, reflection and the coefficients
of the wavefunctions in the bilayer part are determined by imposing the
appropriate boundary conditions at the beginning ($x=0$)  and at
the end ($x=L$) of the bilayer region. Matching of the wavefunctions
amounts to require their continuity. As the Hamiltonian is a
first-order differential equation, current
conservation is ensured automatically. The precise boundary condition depends both on the
lead configuration ($1\rightarrow 1$ or $1\rightarrow 2$) and on the stacking.

\subsubsection{AA stacking}
\label{sec:aa}
In this stacking, each atom $A1(B1)$ has an atom $A2(B2)$ on top of
it. The dispersion in the central part is given by Eq. (\ref{AutoAA}), and 
for each incident carrier with momentum $k_x$, there are always
two reflected and two transmitted eigenfunctions with momenta
$\pm (k_x \pm \gamma_1 / v_F)$; see Fig. \ref{fig:bands}(b).

In the $1\rightarrow 1$ 
(bottom-bottom)
 configuration the wavefunction should be
continuous in the bottom layer, i.e. $ \phi_A^{(1)}(x)$ continuous
at $x=0$ and $\phi_B ^{(1)}(x)$ continuous at $x=L$; for the top
layer 
\begin{equation} \label{cond_3}
\phi_A^{(2)}(x=0)=\phi_B ^{(2)}(x=L) =0 \,  . 
\end{equation}
From these boundary conditions we obtain the transmission
\begin{equation}\label{T_AA_BB}
 T_{AA}^{1 \rightarrow 1}  =
  1 - \frac{\sin ^4  {\frac{\gamma_1  L}{v_F} }}
{1 + 2\cos { \frac{2EL}{v_F} } \cos ^2 { \frac{\gamma _1  L}{v_F} }+ \cos ^4{ \frac{\gamma _1  L}{v_F} } } .
\end{equation}

In the $1\rightarrow 2$  configuration the bottom wavefunction  
$\phi_A^{(1)}(x)$  and the top
wavefunction $\phi_B ^{(2)}(x)$ should be continuous at $x=0$ and $x=L$ respectively. 
In addition, the hard-wall condition should be
satisfied:
\begin{equation}
\phi_A^{(2)}(x=0)=\phi_B ^{(1)}(x=L) =0 \, \, . 
\end{equation}
The above boundary conditions yield the transmission 
\begin{equation}\label{T_AA_BT}
 T_{AA}^{1 \rightarrow 2}  =
 1 - \frac{ \cos ^4 \frac{\gamma _1  L}{v_F} }{2 \left(1 - \cos  \frac{2EL}{v_F} \right)  \sin ^2 \frac{\gamma _1  L}{v_F} + \cos^4 \frac{\gamma _1  L}{v_F}  }  .
\end{equation}
We see from these equations that the conductance changes periodically
as function of the incident energy and length of the bilayer
flake. For fixed $L$, the transmission is a periodic function of the
incident energy. In the bottom-bottom geometry there are
antiresonances, $T_{AA}^{1\rightarrow 1}$=0, at energies given
by $\frac {\pi v_F} L (n+ \frac 1 2)$, with $n=0,1,2...\,$. These
energies corresponds to quasilocalized states in the top part of
the bilayer flake. \cite{com_loc}
The paths through the bottom graphene ribbon and through the
quasilocalized state of the top flake interfere destructively, producing the
antiresonance. \cite{Tekman_1993,Clerk_2001,Wang_2002,Orellana_2003} In the bottom-top configuration, the momenta of the
quasilocalized states of the bilayer flake are shifted in  $- \frac {\pi}{2L}$, so the
antiresonances occur at energies $\frac {\pi v_F} L n $, with
$n=0,1,2... \,$.

For fixed energy,  the conductance varies periodically with the length
of the bilayer flake. There is a period, $\pi v _F /E$,  related
to the energy of the incident carrier; other periods are
harmonics of that imposed by the interlayer hopping, $\pi v_F
/ \gamma _1 $. The dependence of the conductivity on $\gamma _1 $
can be understood by resorting to a simple  non-chiral model with linear dispersion.
Consider an incident carrier from the left with momentum $k_x$ and energy $E=v_F
k_x$ in the bottom sheet. When arriving at the
bilayer central region, the incident wavefunction decomposes into a combination of 
bonding ($b$) and antibonding ($a$) states of the bilayer with
momentum $k^{b(a)}$ = $k_x \pm \gamma _1 / v_F$. The conductance through the bilayer 
region is proportional to the probability of finding an electron at the top (bottom) end of the central region, 
$1\pm \cos (k^b -k^a)L=1 \pm \cos \gamma _1  L /v_F$, depending of whether the system is in the 
$1\rightarrow 2$ or in the $1\rightarrow 1$ configuration. This simple model explains the dependence of the
conductivity on harmonics of $\cos \gamma _1  L / v_F$ and also why 
the $1\rightarrow 2$ and the $1\rightarrow 1$ transmissions are in counterphase. 
The phase opposition is more evident in the $E\rightarrow 0$ limit of Eqs. (\ref{T_AA_BB}) and (\ref{T_AA_BT}), 
which give an  $E=0$ conductance in the bottom-top configuration equal to zero, whereas in the
bottom-bottom configuration it has a maximum finite value that depends on
the flake size:
\begin{equation}\label{T_AA_BB_0}
 T_{AA }^{1 \rightarrow 1}(E=0)  = 1 -\frac{ 4 \sin ^4 \frac{ \gamma _1}{ v_F}  }
 {3 + \cos ^2 \frac{2 \gamma _1  L}{ v _F}},
\end{equation}
\begin{equation}\label{T_AA_BT_0}
 T_{AA}^ {1 \rightarrow 2} (E=0) = 0.
\end{equation}

\subsubsection{AB stacking}
\label{sec:ab}
In this stacking only the atoms $A$ of layer $1$ and the atoms $B$
of layer $2$ are directly connected by tunneling.  The dispersion in
the central part is given by Eq. (\ref{AutoAB_1}). For an incident
carrier with $|E|> \gamma _1$ and momentum $k_x$ there are always two reflected and two transmitted
eigenfunctions with momentum $\pm k _{1 (2)}= \pm \sqrt {k_x ( k_x
\pm \gamma_1 / v_F)}$ in the bilayer region, see Fig. \ref{fig:bands}(c). However, for
incident wavefunctions with $|E|< \gamma _1$,  there are only one
reflected and one transmitted central wavefunctions with momenta $
\pm k _{1} =\pm \sqrt {k_x ( k_x + \gamma_1 / v_F)}$. In addition, 
there are an evanescent and a growing  state  with decay constants 
$\kappa = \pm \sqrt {k_x ( \gamma_1 / v_F- k_x)}$. Therefore, the
conductance  of the system depends on whether the energy of the carrier
is larger or smaller than the interlayer hopping. For $|E|> \gamma
_1$, there are two channels in the central region and the
interference between these channels produces antiresonances, whereas
for $|E|< \gamma _1$ only an electronic channel is present in the
central region, and Fabry-Perot interference can occur. Analytical,
but very large  and impractical expressions can be obtained for the
conductance in the $AB$ stacking.
Therefore, we choose to present
the expressions for the transmission in the low and high energy
limit. In the next section, when comparing with the tight-binding
results, we plot the exact results obtained from wavefunction matching 
in the continuum approximation.

 The boundary conditions for $AB$ stacking in the bottom-bottom configuration are
similar to those of the AA case:  $ \phi_A^{(1)}(x)$ and $\phi_B ^{(1)}(x)$
should be continuous at $x=0$  and $x=L$ respectively,  and
\begin{equation}
\phi_A^{(2)}(x=0)=\phi_B ^{(2)}(x=L) =0 \, \, .
\end{equation}

In the low energy limit, $E \ll \gamma_1$, the $AB$ stacking conductance 
in the bottom-bottom configuration takes the
form
\begin{equation}
\label{T_lim0AB_BB}
T_{AB} ^{1\rightarrow 1} (E \ll \gamma_1)=1-\frac {1}{ 1+\frac
{4 E}{\gamma_1} \frac {( \cos {k_1 L}+ \cosh {\kappa L} ) ^2}{(
\cosh {\kappa L} \sin{k_1 L} -\cos {k_1 L} \sinh {\kappa L} ) ^2}}
\, \, \, ,
\end{equation}
which presents resonances when $\tan{k_1 L } = \tanh {\kappa L}$;  
for large $L$ this occurs when $L= (n + \frac{1}{4})\frac { \pi} {k_1}$, being $n$ an
integer.  For $E \rightarrow 0$ the system has transmission unity.

In the limit of large energy, $E \gg  \gamma_1$ and in the
bottom-bottom configuration the transmission is 
\begin{widetext}
\begin{equation}
\label{T_limgAB_BB}
T_{AB } ^{1 \rightarrow 1} (E \gg \gamma_1)=1-\frac { 8 \sin ^4
\left ( \frac {k_1-k_2} 2 L \right )}{ 11+4 \cos{2k_1 L} + 4
\cos{(k_1-k_2)L}+\cos{2(k_1-k_2)L} + 4 \cos {2 k_2L}+ 8
\cos{(k_1+k_2)L}} \, \,  .
\end{equation}
\end{widetext}
This transmission presents antiresonances associated with
destructive interferences of the two electronic paths in the bilayer
region.  The behavior  of the conductance is similar to that of
the $AA$ stacking, Eq. (\ref{T_AA_BB}). There are periodicities
associated with the energy of the incident electron: for $E \gg
\gamma _1$ , $2k_1 L \sim 2k_2 L \sim (k_1 + k_2) L \sim 2E L/v_F$;  and there are also 
periodicities associated with the interlayer hopping. The lower
harmonic in the $AB$ stacking, $  \frac {k_1-k_2} 2 L \sim
\frac{\gamma_1 L} {2v_F}$, is half the basic harmonic in the $AA$ stacking,
and this reflects the fact that in the $AB$ stacking only half of
the atoms have direct interlayer tunneling.

In the bottom-top geometry $ \phi_A^{(2)}(x)$ and $ \phi_B^{(1)}(x)$
should be continuous at $x=0$ and $x=L$ respectively, and
\begin{equation}\label{order}
\phi_B^{(2)}(x=0)=\phi_A ^{(1)}(x=L) =0 \, \, \, .
\end{equation}
In the $AB$ stacking, interlayer tunneling connects  $A1$ atoms
with $B2$ atoms; this arrangement 
determines the form of
Eq. (\ref{order}).

For $E<< \gamma_1$ the bottom-top transmission can be approximated
as
\begin{widetext}
\begin{equation}
\label{T_lim0AB_BT}
T_{AB} ^{1 \rightarrow 2} (E \ll  \gamma_1)=1- \frac { \left (
1+\cos {k_1 L}\cosh{\kappa L} + \frac E {\gamma_1} \sin {k_1 L}
\sinh {\kappa L} \right ) ^2} {4 \frac E {\gamma_1} \left ( \cosh
\kappa L \sin k_1 L + \cos k_1 L \sinh {\kappa L} \right ) ^2 +
\left ( 1+ \cos{k_1 L} \cosh{\kappa L} - 3 \frac E {\gamma_1} \sin
{k_1 L} \sinh {\kappa L} \right ) ^2}.
\end{equation}
\end{widetext}
It can be seen that the $1 \rightarrow 2 $ conductance goes to zero when 
For $E \rightarrow 0$, $T_{AB} ^{1\rightarrow 2}$ tends to zero, and it is complementary to 
 $T_{AB} ^{1\rightarrow 1}$. For large $L$, $T_{AB} ^{1\rightarrow 2}$ presents resonances at 
 $L= (n + \frac{1}{2})\frac { \pi} {k_1}$, with $n$ integer. 
For energies larger than the interlayer hopping the conductance can
be approximated as 
\begin{widetext}
\begin{equation}
\label{T_limgAB_BT}
T_{AB } ^{1 \rightarrow 2} (E \gg \gamma_1)=1-\frac { 8 \cos ^4
\left ( \frac {k_1-k_2} 2 L \right )}{ 11-4 \cos{2k_1 L} - 4
\cos{(k_1-k_2)L}+\cos{2(k_1-k_2)L} - 4 \cos {2 k_2L}+ 8
\cos{(k_1+k_2)L}} \, \, \, .
\end{equation}
\end{widetext}
In this energy limit, the interference between different electron paths through the systems produces antiresonances. 
Similarly to the bottom-bottom configuration, for a fixed energy $E \gg \gamma_1$ the transmission varies periodically with $L$,
with one period given by the incident energy, $\pi v_F/E$, and others related to the interlayer hopping, $\propto \pi v_F/\gamma_1$.

\section{\label{sec:results}  Results}

As the systems possess electron-hole symmetry, we concentrate on energies $E \geq 0$.
Let us recall here that the length of a unit cell (u.c.) for an armchair ribbon is $3a_{CC}=\sqrt{3}a_0$.  In the following Figures,
we choose to give the system length $L$  in terms of the armchair ribbon u.c. length, which is unambiguous for the discrete tight-binding model. Note that, in the continuum approximation, the hard wall conditions at the edges of the system ($x=0$ and $x=L$) are set at two 
extra rows of atoms where the wavefunctions are imposed to be zero. This amounts to add to the system length the quantity $a_{CC}$, which we take into account when comparing the continuum and the tight-binding results. 

\subsection{$AA$ and $AB$-$\alpha$ stackings}

\begin{figure}[ht]
\includegraphics[width=\columnwidth,clip]{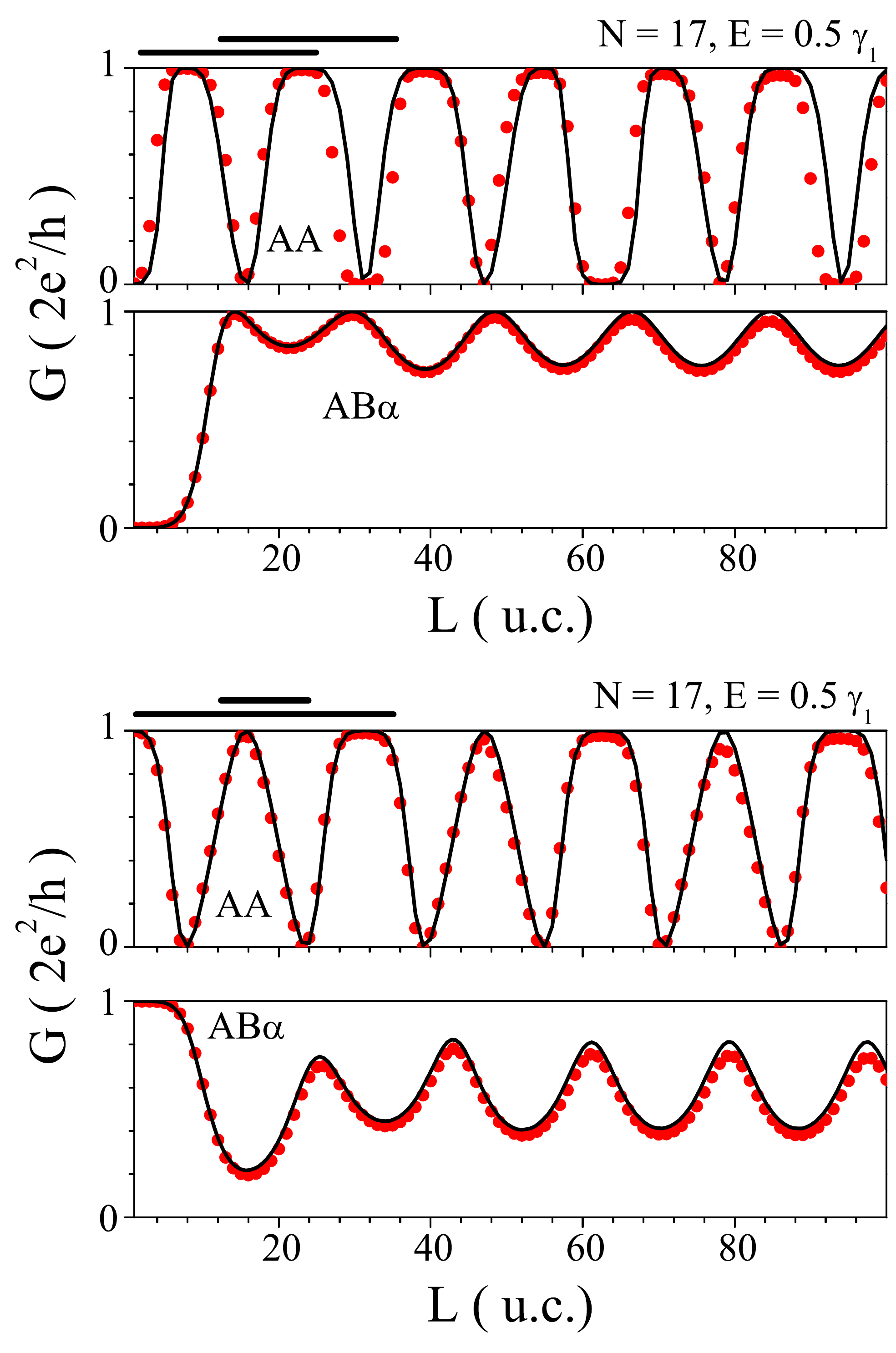}
\caption{(color online) Conductance as a function of the bilayer region length $L$  for a ribbon of width N = 17 with
$AA$ and $AB$-$\alpha$ stackings, at a Fermi energy $E = \gamma_1/2$. The top panel shows  the $ 1 \rightarrow 2$ configuration
and the lower panels are for the $ 1 \rightarrow 1$ configuration, as schematically indicated in the upper left corners of the panels. The plots are labeled with the stackings ($AA$ and $AB$-$\alpha$). Red circles:  tight-binding results. Black solid lines:  continuum model calculations.
}
\label{fig:gvslh}
\end{figure}

As discussed in the previous Section, the expressions for the transmission (Eqs. (\ref{T_AA_BB}), (\ref{T_AA_BT}),  (\ref{T_lim0AB_BB}),  (\ref{T_lim0AB_BT}), (\ref{T_limgAB_BB}), and (\ref{T_limgAB_BT})) demonstrate that the dependence with the system length
has periodicities related to the interlayer coupling $\gamma_1$. This is evident in Fig. \ref{fig:gvslh}, which shows the length dependence of the
conductance at a fixed energy $E=  \gamma_1/2$, for the stackings $AA$ and AB-$\alpha$ and the two lead configurations, $ 1 \rightarrow 2$ and $ 1 \rightarrow 1$. Here we depict the tight-binding results with circles and the continuum ones with full lines. The tight-binding calculations are performed  for a ribbon of width $N=17$, but for this energy range only one channel contributes to the conductance in the monolayer and at most two channels in the bilayer flake, so the conductance is independent of $N$. 
The agreement between the two models is excellent for these stackings and energy range. As expected, the $AA$ stacking shows clear antiresonances as a function of length, and the results for the $ 1 \rightarrow 2$ and $ 1 \rightarrow 1$ configurations are exactly in counterphase.  

The results for the two configurations ($ 1 \rightarrow 1$ and $ 1 \rightarrow 2$) with $AB$-$\alpha$ stacking have an approximate complementarity; only at $L \rightarrow 0$ there is  zero transmission for the $ 1 \rightarrow 2$ case corresponding to a transmission maximum for the $ 1 \rightarrow 1$ system. The subsequent maxima and minima are slightly shifted, and more importantly, there are no zero antiresonances for finite length. As mentioned before, there is only one transmission channel in the bilayer, so although the conductance oscillates due to finite-size effects, there are not antiresonances for the $AB$-$\alpha$ at this energy.

\begin{figure}[ht]
\includegraphics[width=\columnwidth,clip]{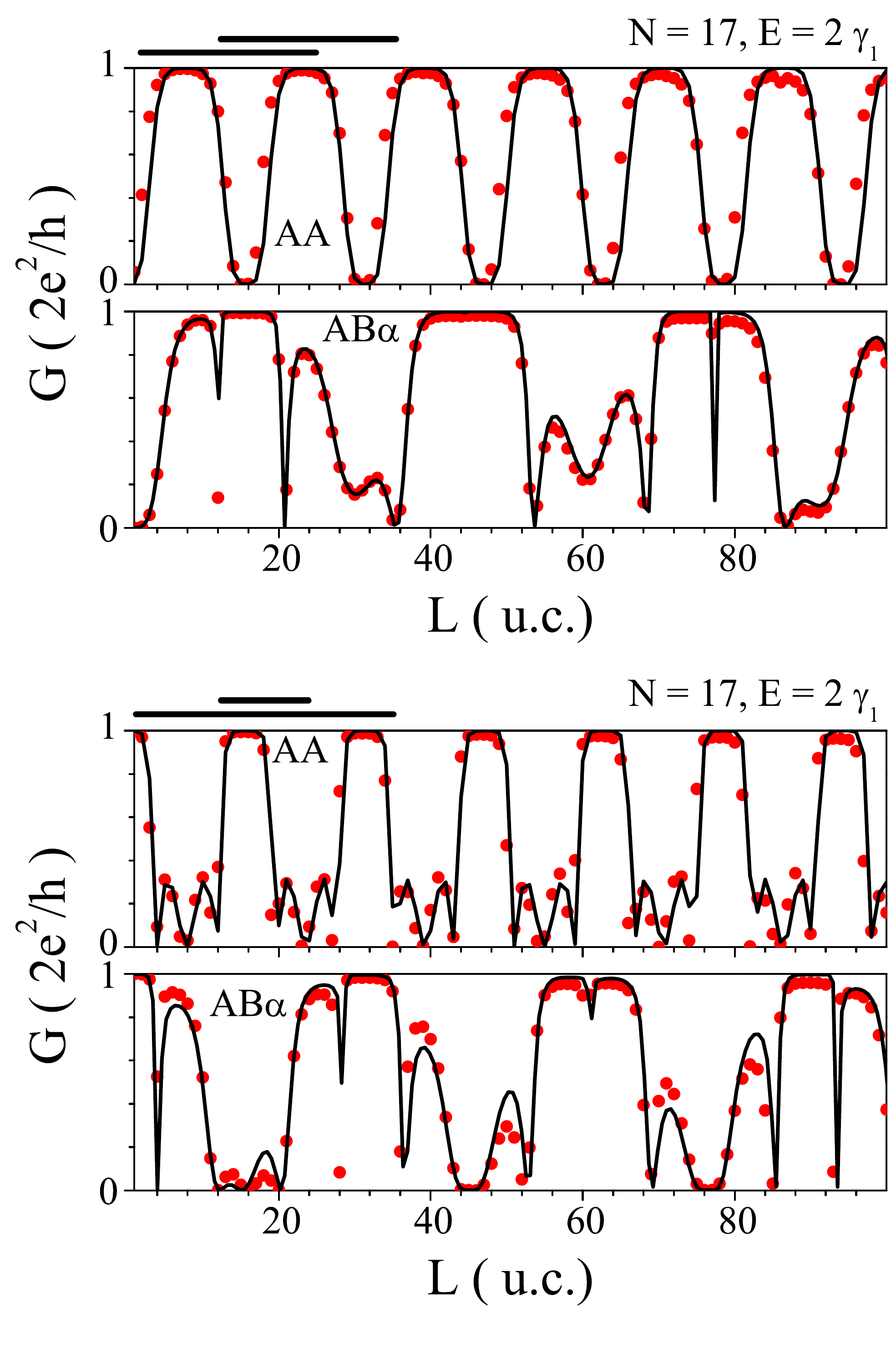}
\caption{
(color online)
Conductance as a function of the bilayer region length $L$  for a ribbon of width N = 17 with
direct and Bernal-$\alpha$ stackings, at a Fermi energy $E = 2  \gamma_1$. The top panel shows  the $ 1 \rightarrow 2$ configuration
and the lower panels are for the $ 1 \rightarrow 1$ configuration, as schematically indicated in the upper left corners of the panels.The plots are labeled with the stackings ($AA$ and $AB$-$\alpha$). Red circles:  tight-binding results. Black solid lines:  continuum model calculations.
}
\label{fig:gvsld}
\end{figure}

Fig. \ref{fig:gvsld} shows the length dependence of the
conductance for another energy $E= 2  \gamma_1$, where there are two conducting channels for both stackings. It is apparent the change for the
AB-$\alpha$ case, which now presents antiresonances with zero conductance. As to the $AA$ stacking, the conductance for the $ 1 \rightarrow 2$ configuration shows only one clear period of 16 u.c., whereas the $ 1 \rightarrow 1$ case shows also a 8 u.c. periodicity, stemming from the $\cos 2 E L/v_F$ term in the conductance. The analytical expressions Eqs. (\ref{T_AA_BB}) and (\ref{T_AA_BT}) allows us to verify that, for the $ 1 \rightarrow 2$ case, this energy-dependent term  $\cos 4 \gamma_1 L  /v_F$ combines with the other $\gamma_1 $-dependent terms to yield a single period, whereas for the $ 1 \rightarrow 1$ case, the $\cos 4 \gamma_1 L/v_F $
survives. 

\begin{figure}[ht]
\includegraphics[width=\columnwidth,clip]{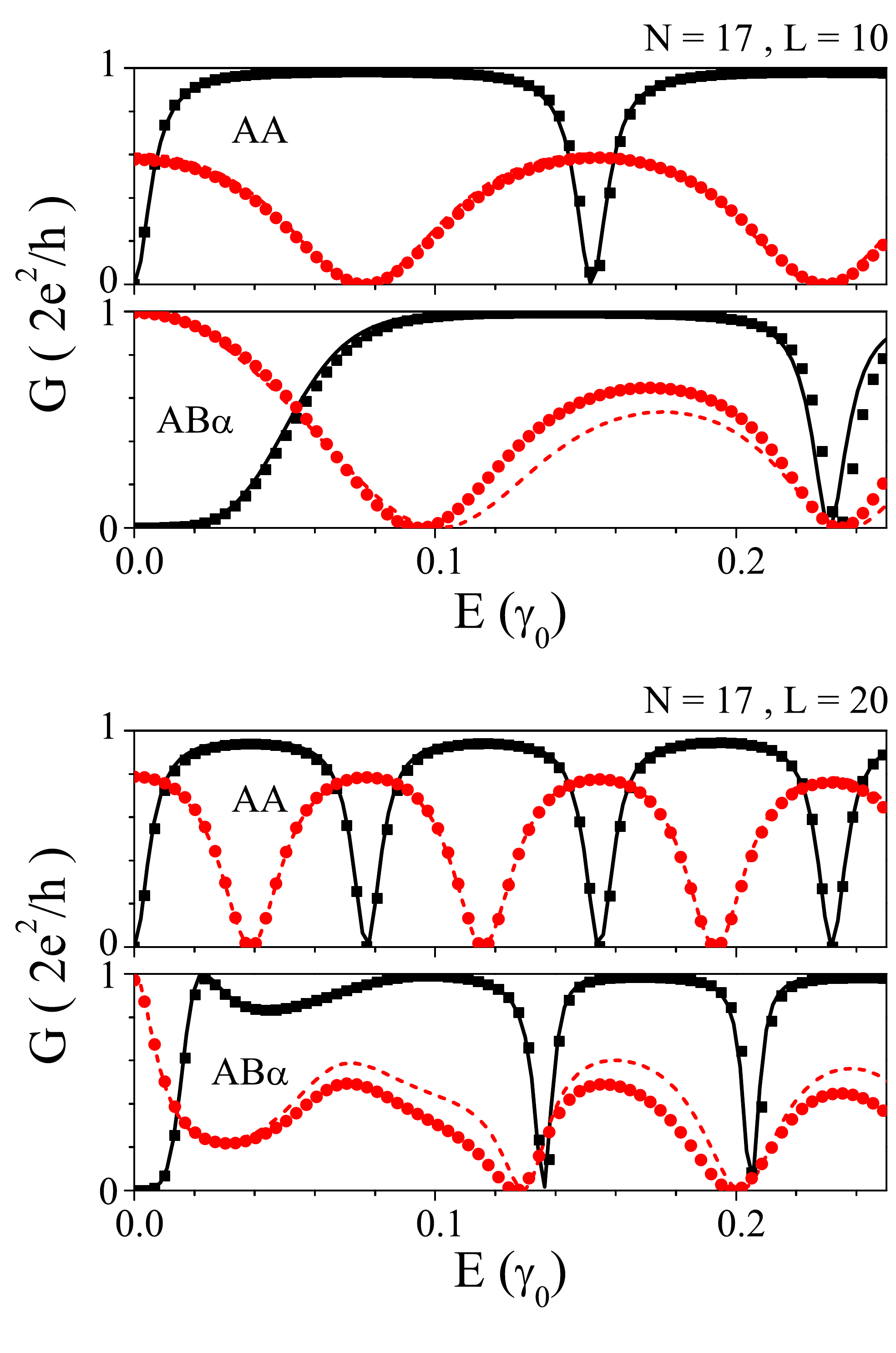}
\caption{(color online)
Conductance as a function of Fermi energy for  a bilayer region of length L = 10 u.c. (top panel) and L = 20 u.c. (bottom panel). Dirac-like results: solid black lines  correspond to the $ 1 \rightarrow 2$ configuration
and the dashed red lines  for the $ 1 \rightarrow 1$ configuration.  Inside each panel, the top graph depicts the $AA$ stacking, and the bottom graph shows the $AB$-$\alpha$ stacking data, as labeled therein. The tight-binding calculations are shown in colored circles. 
}
\label{fig:gvse17}
\end{figure}

Figure \ref{fig:gvse17} shows the conductance $G(E)$ as a function of energy for the two geometries considered and the most symmetric stackings, namely the $AA$ and the $AB$-$\alpha$, for a flake length of $L=10$ u.c. (top panel) and  $L=20$ u.c. (bottom panel). The tight-binding results are depicted with circles and the continuum ones with lines.

As discussed before, the most characteristic feature of the transmission is the appearance of Fano antiresonances with zero conductance.
This can happen for any energy in the case of $AA$ stacking because there are always two conducting channels in the $AA$ bilayer. On the contrary, for $AB$-$\alpha$ stacking, with only one channel for $E <  \gamma_1 $, the oscillations in the conductance are due to a Fabry-Perot-like effect, i.e., the interference of one scattering channel with itself due to the finite size of the structure. This is most clearly seen for the
$L=20$ case, where the $AB$-$\alpha$ stacking presents a non-zero minimum in the conductance in the $(0,  \gamma_1)$ energy range, whereas the antiresonances above  $ \gamma_1$ clearly reach zero values.

Notice as well the agreement with the continuum calculations when $E \rightarrow 0$. All the $ 1 \rightarrow 2$ configurations have zero conductance in this limit. As to the $ 1 \rightarrow 1$ configuration, $G(E=0)$ has a in general  nonzero value, which oscillates as a function of the system length, as described by
Eq. (\ref{T_AA_BB_0}) and it is seen in Fig. \ref{fig:gvse17}.
 The $ 1 \rightarrow 1$ and $ 1 \rightarrow 2$ results for the $AA$ stacking do show a certain complementarity: the conductance minima in one configuration coincide with the maxima of the other one.
Furthermore, the periodicity of the conductance as a function of energy for the $AA$ stacking  due to the term $\cos 2EL/v_F$ is evident in Fig. \ref{fig:gvse17}, and agrees perfectly with the value given by the Dirac continuum approximation, namely  0.16 $\gamma_0$ for $L=10$ and
0.079 $\gamma_0$ for $L=20$.

\begin{figure}[ht]
\includegraphics[width=\columnwidth,clip]{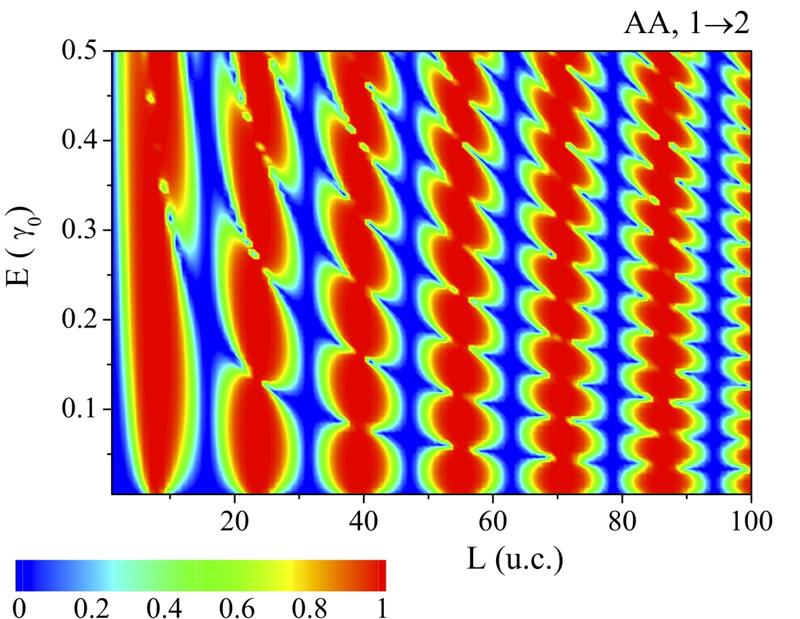}
\includegraphics[width=\columnwidth,clip]{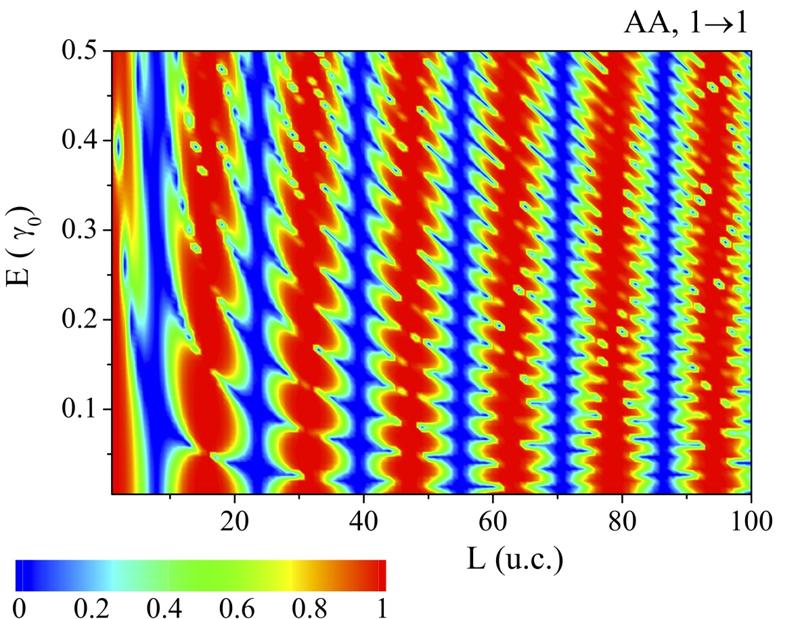}
\caption{
(color online)
Transmission as a function of the energy and flake length for the $AA$ stacking, as obtained from the continuum Dirac model. Top panel: $ 1 \rightarrow 2$ configuration. Bottom panel: $ 1 \rightarrow 1$ configuration.
}
\label{fig:contaa}
\end{figure}

The continuum model allows us to make a more complete characterization of the behavior of these systems. Fig. \ref{fig:contaa} shows contour plots of the transmission versus energy and bilayer flake length for the direct stacking and the two configurations, $ 1 \rightarrow 1$ and $ 1 \rightarrow 2$,  given by Eqs. (\ref{T_AA_BB}) and (\ref{T_AA_BT}).
We clearly see the main transmission antiresonances with a 16 u.c. period, 
stemming from the interlayer hopping term $\pi v_F/\gamma_1$, as discussed in Sec. \ref{sec:aa}. 
In fact, it turns out that for certain flake sizes $L$, the conductance is
zero, independently of the energy. As this spatial period depends directly on the interlayer coupling strength $\gamma_1$, we propose that this feature can be used to measure the interlayer hopping parameter, the value of which is still under debate:\cite{Nilsson_2006} by overlapping two nanoribbons and displacing one of them with respect to the other, the spatial period could be measured and thus $\gamma_1$ would be obtained. 

As the variation of the conductance as a function of length is so dramatic, from one quantum of conductance to zero, this system can function as an electromechanical switch, turning from the maximum conductance to zero by a displacement of a few \AA .
The contour plots for the $AA$ case also reveal the counterphase behavior of the $ 1 \rightarrow 1$ and $ 1 \rightarrow 2$ configuration discussed previously. It is patent how the maxima of the conductance vs. L for the $ 1 \rightarrow 2$ system coincide with the $ 1 \rightarrow 1$ minima and viceversa.

\begin{figure}[ht]
\includegraphics[width=\columnwidth,clip]{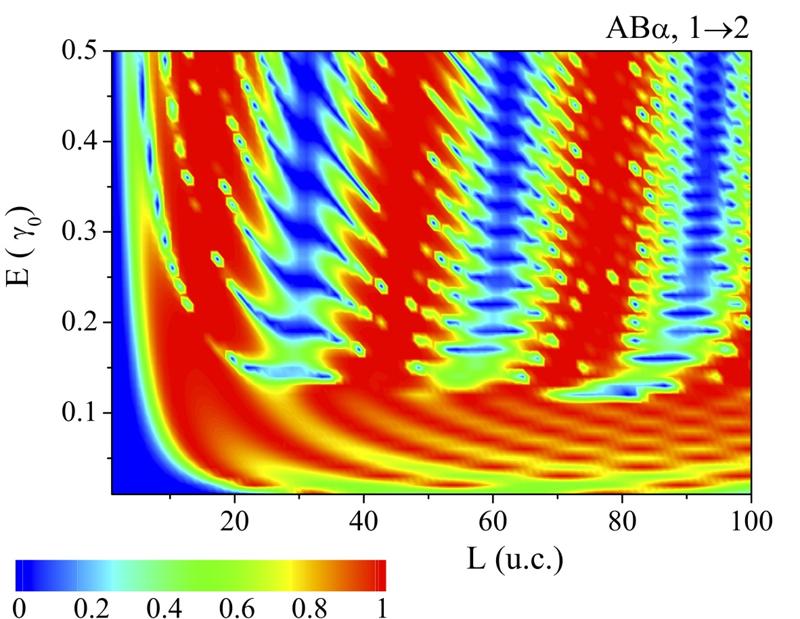}
\includegraphics[width=\columnwidth,clip]{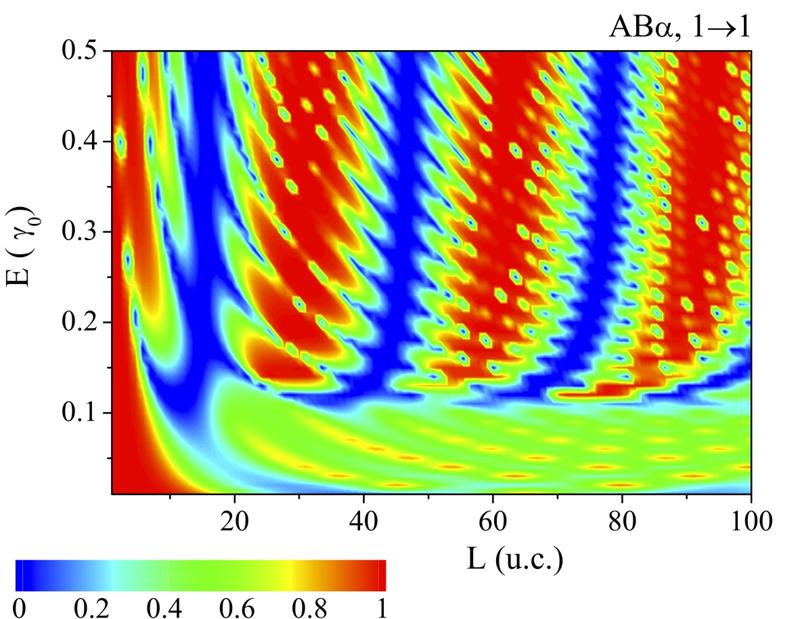}
\caption{
(color online)
Transmission as a function of the energy and flake length for the $AB$-$\alpha$ stacking, as obtained from the continuum Dirac model. Top panel: $ 1 \rightarrow 2$ configuration. Bottom panel: $ 1 \rightarrow 1$ configuration.
}
\label{fig:contab}
\end{figure}

Fig. \ref{fig:contab} displays the contour plots for the $AB$-$\alpha$ case obtained within the Dirac model. There are two important differences with respect to the $AA$ stacking. First, now there are two distinct energy regions, set by the interlayer coupling $\gamma_1$. Below $E=\gamma_1$, there are no antiresonances because there is only one propagating channel at the bilayer. There are conductance oscillations, but not so marked as for $E>\gamma_1$, where the zero antiresonances appear because of the coexistence of two propagating eigenchannels in the bilayer flake. For this energy range, the behavior is more similar to that found for the $AA$ stacking, with an obvious difference on the spatial periods.  As already mentioned in Sec. \ref{sec:ab}, the lower
harmonic in the $AB$ stacking is $\frac{\gamma_1 L} {2v_F}$, thus yielding a longer spatial period (32 u.c.) that we attribute to the smaller coupling between layers for this case.

\subsection{$AB$-$\beta$ stacking}

Until this point, we have focused in the more symmetric stackings, for which the continuum Dirac model and the tight-binding have an excellent agreement, as demonstrated. Now we turn our attention to the $AB$-$\beta$ stacking, which we can only model adequately with
the tight-binding approach. This is because of the lack of symmetry of the ribbon edges, as it can be seen in Fig. \ref{fig:bands} (d). The atoms at the upper egde of the top layer are not connected to the atoms of the bottom layer, independently of the sublattice they belong, and viceversa.
Such a feature cannot be well described by the continuum Dirac Hamiltonian given by Eq. (\ref{HAB}), which assumes that all carbon atoms in the $A$ sublattice of the bottom layer are connected to the  $B$ atoms on the top layer. This difference is not very important for wide ribbons, but it is noticeable for the narrow cases, for which the proportion of atoms at the ribbon edges is non-negligible.

\begin{figure}[ht]
\includegraphics[width=\columnwidth,clip]{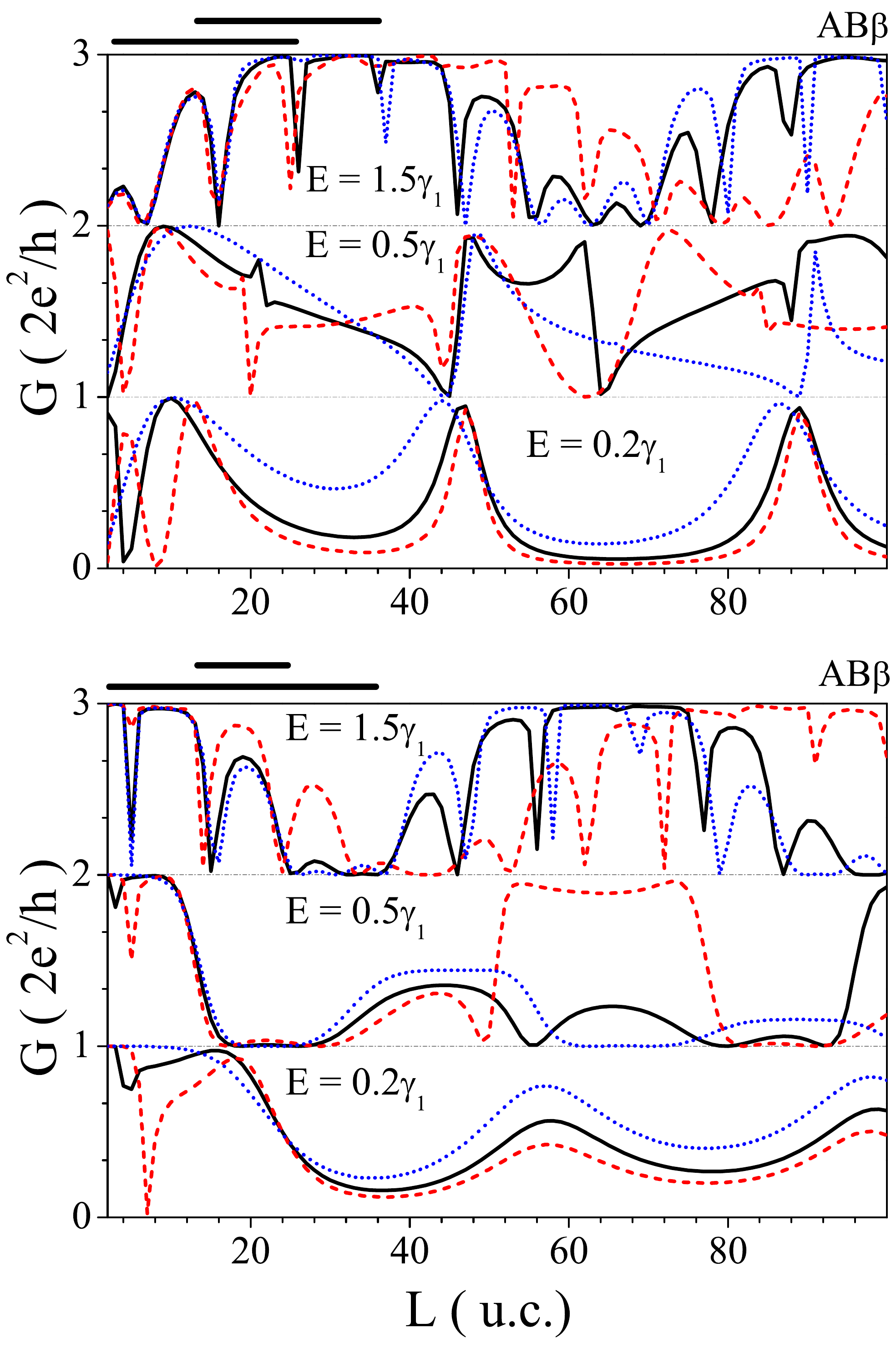}
\caption{
(color online)
Conductance as a function of the length of bilayer region in AB-$\beta$ stacking for three Fermi energies ($E = 0.2 \gamma_1$, $E = 0.5 \gamma_1$ and $E = 1.5 \gamma_1$)
for three ribbon widths: $N = 5$ (dotted blue line),  $N = 17$ (black solid line), and $N = 29$ (red dashed line). The $E = 0.5  \gamma_1$ and $E = 1.5 \gamma_1$ curves have been shifted up in one and two conductance units respectively for the sake of clarity.
Top panel: $ 1 \rightarrow 2$ configuration. Bottom panel:  $ 1 \rightarrow 1$ configuration.  }
\label{fig:bernal}
\end{figure}

One way to assess the importance of the edge effect is to check the energy difference between the first and the second subband for $E\geq 0$. For a $AB$-$\alpha$ nanoribbon is always $\gamma_1$, whereas for $AB$-$\beta$ nanoribbons it depends on the ribbon width, as it can be seen in  Figs. \ref{fig:bands} (c) and (d). Size effects are related to the ratio of atoms which are not well described by the continuum $AB$ Hamiltonian of Eq. (\ref{HAB}). 
This brings in a dependence on the ribbon
width, as shown in  Fig. \ref{fig:bernal}, depicting the conductance for three energies and ribbon widths $N$ for the two configurations, $ 1 \rightarrow 1$ and $ 1 \rightarrow 2$. Notice the dependence on the ribbon width; the conductance results demonstrate that size effects are still important for $N\approx 30$. 
For the lowest energy depicted, for which there is only one propagating channel in the bilayer flake, the three widths show a similar behavior for sufficiently long  flakes ($L> 10$). However, for the highest energies the disagreement is patent, due to the dependence of the longest spatial period on the system width. The different periods are more clear for the energy $E=1.5 \gamma_1$, for which at least half a wavelength of the oscillation can be appreciated for the three ribbon widths. Notice that the case $E=0.5 \gamma_1$, shown in the central part of both panels in Fig. \ref{fig:bernal}, is also depicted for the $AB$-$\alpha$ stacking in Fig. \ref{fig:gvslh}. This striking difference in the conductance for the two $AB$ stackings is due to the fact that in the $AB$-$\alpha$ case theres is only one channel for this energy, whereas in the $AB$-$\beta$ there are already two. 

\section{Summary}

In this work, we have studied the conductance of a graphene bilayer flake contacted by two monolayer nanoribbons. Two contact geometries have been considered: either the left and right lead are contacted to the same layer of the flake or to opposite layers. Furthermore,  three different stackings for the graphene flake have been taken into account, namely, $AA$, $AB$-$\alpha$ and $AB$-$\beta$. 

 We have calculated the conductance with a tight-binding approach and also by performing a mode-matching calculation within the continuum Dirac model, by choosing the appropriate boundary conditions.
We have explained the features in the transmission and obtained analytical expressions that allow us to elucidate the transport characteristics of these structures. We have found several periodicities on the conductance, related to the energy and the interlayer coupling of the system. 

In particular, for the $AA$ configuration, we have found that the conductance through the flake shows Fano antiresonances, that we have related to the interference of two different propagating channels in the structure. For a flake of length $L$, the main transmission period is given by $\pi v_F/L$. For a fixed incident energy, the conductance as a function of the system length $L$ oscillates with two main periods related to the energy $E$ and the interlayer coupling $\gamma_1$.

 For the $AB$ stacking, we have found two distinct behaviors as a function of the incident energy $E$: for energies larger than the interlayer hopping $\gamma_1$, the transmissions resemble those found for the $AA$ stacking. This is due to the existence of two propagating channels at this energy range. There is, however, a difference on the main period related to the interlayer hopping $\gamma_1$, which is twice the period found for the $AA$ stacking. This can be understood by noticing that in the $AB$ stacking only half of the atoms are connected between the two graphene layers. For energies smaller than $\gamma_1$, the $AB$-stacked flake only has one eigenchannel, and the conductance shows resonances related to the existence of Fabry-Parot-like states in the system.

The conductance of these bilayer flakes can oscillate between zero and the maximum conductance as a function of length; thus, a system composed by two overlapping nanoribbons can behave as an electromechanical switch. 
We propose that these characteristics can be employed to measure the interlayer hopping in bilayer graphene. Our results constitute a comprehensive view of transport through bilayer graphene flakes, clarifying the role of stacking, contact geometries, flake width and length in the conductance of these structures.


\section*{Acknowledgments}
This work has been partially supported by the Spanish DGES under
grants MAT2006-06242, MAT2006-03741, FIS2009-08744 and Spanish CSIC under grant
PI 200860I048.
J.W.G. would like to gratefully acknowledge helpful discussion to Dr. L. Rosales, to the ICMM-CSIC
for their hospitality and MESEUP research internship program. J.W.G. and M.P. acknowledge the financial support of CONICYT/Programa
Bicentenario de Ciencia y Tecnolog\'{\i}a (CENAVA, grant ACT27) and USM 110856 internal grant.

\end{document}